\documentstyle[twoside,fleqn,espcrc2,epsf]{article}

\newcommand{\be}{\begin{equation}}
\newcommand{\ee}{\end{equation}}

\newcommand{\AmS}{{\protect\the\textfont2
  A\kern-.1667em\lower.5ex\hbox{M}\kern-.125emS}}

\title{ Perturbative Matching of Heavy-Light Currents with 
NRQCD Heavy Quarks }

\author{
        Colin J. Morningstar 
        \address{Physics Department, University of California at San Diego,
        La Jolla, CA 92093, USA.}
        and
        Junko Shigemitsu
        \address{Physics Department, 
        The Ohio State University, 
        Columbus, OH 43210, USA.}
                           }
       
\begin{document}

\begin{abstract}

We present further results for one-loop matching of heavy-light 
axial and vector currents between continuum QCD and a lattice theory 
with NRQCD heavy quarks and massless clover quarks.

\end{abstract}

\maketitle

\section{Introduction}

\noindent
Lattice studies of hadronic matrix elements require matching 
between operators in full continuum QCD and those in the 
lattice theory being simulated. In the present article we will focus 
on an effective lattice theory which combines
 nonrelativistic (NRQCD) heavy quarks and clover light quarks.
We wish to match the theories correct through 
$O({p \over M} \, ,\, \alpha { p \over M} \, , \, 
\alpha \, ap)$.

For heavy meson (e.g. $B$ and $B^*$)  leptonic and semileptonic 
decays, the relevant operators are heavy-light vector and axial vector 
currents, denoted in the continuum theory as
$V_\mu \equiv \overline{q} \gamma_\mu h$ and 
$A_\mu \equiv \overline{q} \gamma_5 \gamma_\mu h$.  The first question 
that arises concerns the number and type of  operators required in the 
effective theory.  

\vspace{.1in}
\noindent
For \framebox{ $A_0$ and  $V_0$} one finds 
3 operators in the effective theory through 
 $O(\frac{p}{M})$

\vspace{.1in}
  $J^{(0)}_t = \bar q(x) \, \Gamma_t 
\, Q(x), \qquad \qquad$ 
 ($\Gamma_t = \gamma_5 \gamma_0$ or $ \gamma_0$)

\vspace{.1in}
$ J^{(1)}_t =  \frac{-1}{2M} \bar q(x)
    \,\Gamma_t\,\mbox{\boldmath$\gamma\!\cdot\!\nabla$} \, Q(x), $ 

$ J^{(2)}_t =  \frac{-1}{2M} \bar q(x)
    \,\mbox{\boldmath$\gamma\!\cdot\!\overleftarrow{\nabla}$}
    \,\gamma_0\Gamma_t\, Q(x) $,

\vspace{.2in}
\noindent
plus a dimension 4  discretization correction at 
 $O(\alpha \, ap)$ :

$ J^{disc}_t =  -a \, \bar q(x)
    \,\mbox{\boldmath$\gamma\!\cdot\!\overleftarrow{\nabla}$}
    \,\gamma_0\Gamma_t\, Q(x) $  $=$  $2 \, aM \, J^{(2)}
_t $ 

\vspace{.1in}
 $J^{(0)}_t \longrightarrow J^{(0),imp}_t 
= J^{(0)}_t +  (\alpha \, \zeta_{disc}^{A_0/V_0}) J^{disc}_t 
 \qquad  $

\noindent
(the spinors $h$ and $Q$ are related via a Foldy-Wouthuysen transformation).

\vspace{.1in}
\noindent
For \framebox{ $A_k$ and  $V_k$} one has
5 operators in the effective theory through 
 $O(\frac{p}{M})$

\vspace{.1in}
 $J^{(0)}_k = \bar q(x) \, \Gamma_k 
\, Q(x), \quad \qquad$   
($\Gamma_k = \gamma_5 \gamma_k$ or $\gamma_k$)

\vspace{.1in}
$ J^{(1)}_k =  \frac{-1}{2M} \bar q(x)
    \,\Gamma_k\,\mbox{\boldmath$\gamma\!\cdot\!\nabla$} \, Q(x),$ 

\vspace{.1in}
$ J^{(2)}_k =  \frac{-1}{2M} \bar q(x)
    \,\mbox{\boldmath$\gamma\!\cdot\!\overleftarrow{\nabla}$}
    \,\gamma_0\Gamma_k\, Q(x), $ 

\vspace{.1in}
$ J^{(3)}_k =  \frac{-1}{2M} \bar q(x)\, \Gamma_s \nabla_k 
\, Q(x),  \qquad$
($ \Gamma_s = \gamma_5$ or $\hat{I}$)

\vspace{.1in}
$ J^{(4)}_k =  \frac{1}{2M} \bar q(x)
    \,\overleftarrow{\nabla}_k  \Gamma_s \, Q(x) $.

\vspace{.3in}
\noindent
Again there is an $O(\alpha \, ap)$ discretization correction 
to $J^{(0)}_k$.
\vspace{.1in}
 $J^{disc}_k = 2\, aM \, J^{(2)}_k$ and 

 $ J^{(0)}_k \longrightarrow J^{(0),imp}_k 
= J^{(0)}_k +  (\alpha \, \zeta_{disc}^{A_k/V_k})
 J^{disc}_k  $

\vspace{.2in}
\noindent
The desired relation between  QCD hadronic matrix elements and 
 nonperturbatively determined lattice current matrix elements is hence,
\begin{eqnarray}
 \langle\, A_0\, \rangle_{QCD} &=& \sum_{j=0}^2 
 C^{A_0}_j \, \langle\, J^{(j)}_{A_0} \,\rangle_{LAT},
\qquad \qquad 
\nonumber \\
 \langle\, V_k \,\rangle_{QCD} &=& \sum_{j=0}^4 
 C^{V_k}_j \, \langle\, J^{(j)}_{V_k} \,\rangle_{LAT}, 
\qquad \qquad   \nonumber
\end{eqnarray}
and similarly for  $V_0$ and  $A_k$.  

\vspace{.1in}
\noindent
The matching coefficients, $C_j$, have the following 
perturbative expansion, 
 
\vspace{0.5in}

$ C_j =  1 + \alpha \;  \rho_j  + O(\alpha^2)
\qquad   j = 0,1 $ 

$  C_j =  \alpha \;  \rho_j  + O(\alpha^2) 
\qquad \qquad  j > 1$

\vspace{.1in}
\noindent
The goal is to calculate the  $\rho_j$'s.

\section{ Some Calculational Details}

In the continuum theory we employ on-shell renormalization with naive 
dimensional regularization and the $\overline{MS}$ scheme.  A gluon mass 
is introduced to handle IR divergences that eventually cancel between 
continuum and the lattice.  The light quark mass is set equal to zero.  
On the lattice we worked mainly with an NRQCD action 
correct through $O(p/M)$.  Results with higher order relativistic 
corrections also exist.  The lattice light quarks are massless 
clover fermions.  The light and heavy quark actions and the lattice 
current operators are all tadpole improved.

\section{ Results for One-Loop Coefficients}

In figures 1. and 2. we plot the one-loop coefficients 
$\rho_0$ and $\rho_j/(2aM)$, $j > 0$, for all four current types, 
$V_k$, $A_k$, $A_0$, and $V_0$ versus the inverse dimensionless bare 
heavy quark mass \cite{pert1}.
 The ``bursts'' show $\rho_0$ with 
$log(aM)$ fixed to $log(2)$, where the choice of $aM = 2$ is 
somewhat arbitary, corresponding to the value of the bare 
dimensionless heavy quark mass appropriate for the physical $b$-quark 
on $a^{-1} \approx 2$GeV lattices.  With the logarithmic dependence taken out 
one can compare with $\rho_0$ for the static theory, as indicated 
for $V_k$ and $A_0$\cite{bp}.
  One also sees that $\rho_2/(2aM)$ (the ``diamonds'') goes to 
a non-vanishing  value as $aM \rightarrow \infty$.  This is due to 
the mixing with the discretization correction, $J_\mu^{disc}$. 

\begin{figure}[t]
\begin{center}
\leavevmode
\epsfxsize=8.5cm
\epsfbox[60 120 530 430]{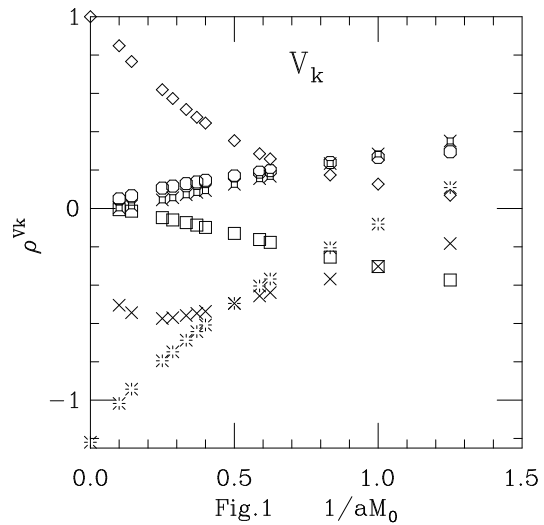}
\end{center}
\caption{ One-loop Coefficients $\protect\rho_j$. 
crosses : $\protect\rho_0$ 
squares : $\protect\rho_1/2aM$ ;
diamonds : $\protect\rho_2/2aM$ (includes $\protect\zeta_{disc}$) ;
octagons : $\protect\rho_3/2aM$ ;
fancy squares : $\protect\rho_4/2aM$ ;
bursts : $\protect\rho_0$ with $log(aM)$ fixed at $log(2)$.
}
\end{figure}

\begin{figure}[t]
\begin{center}
\leavevmode
\epsfxsize=8.5cm
\epsfbox[60 120 530 430]{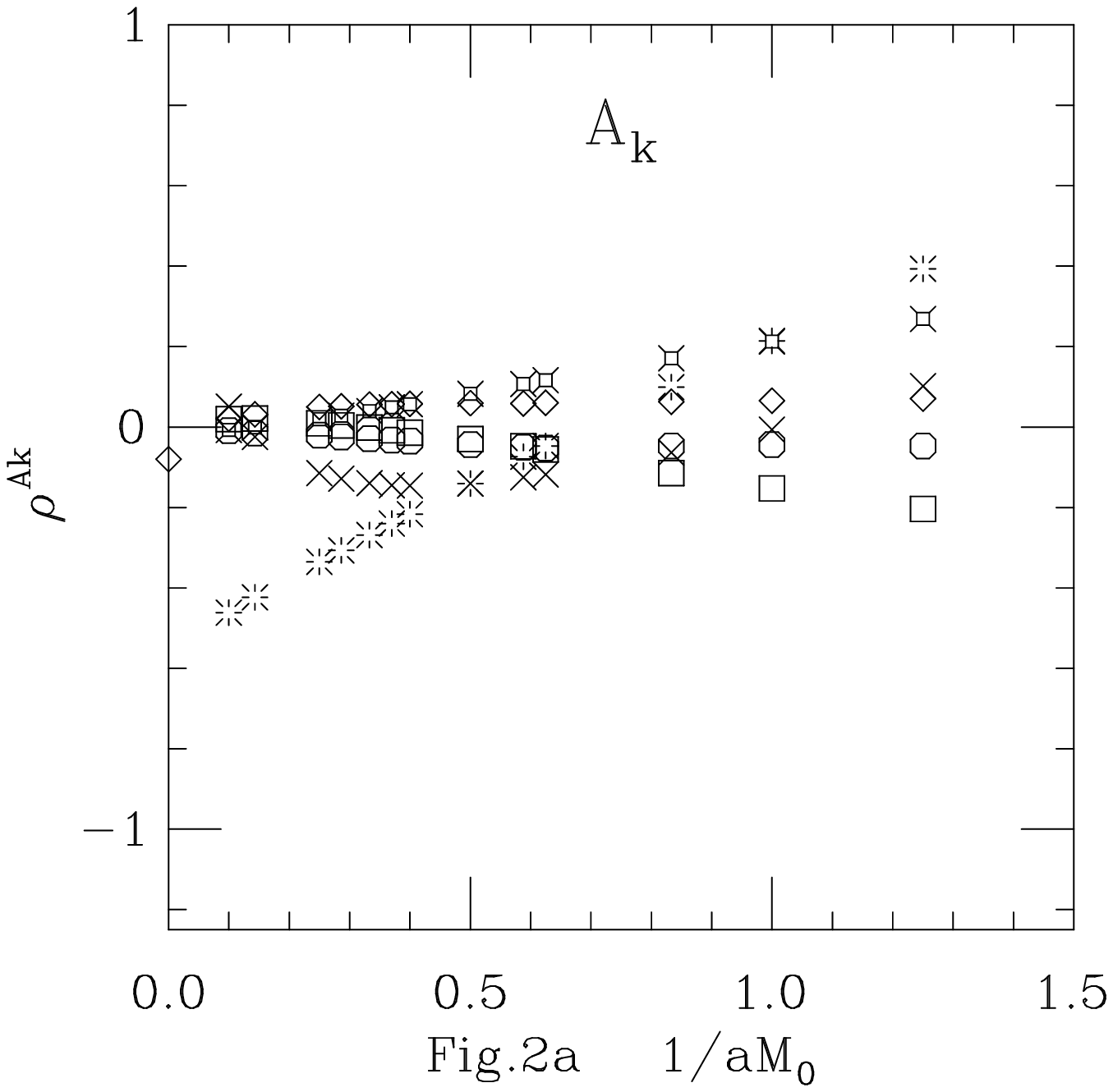}\\[25mm]
\epsfxsize=8.5cm
\epsfbox[60 120 530 430]{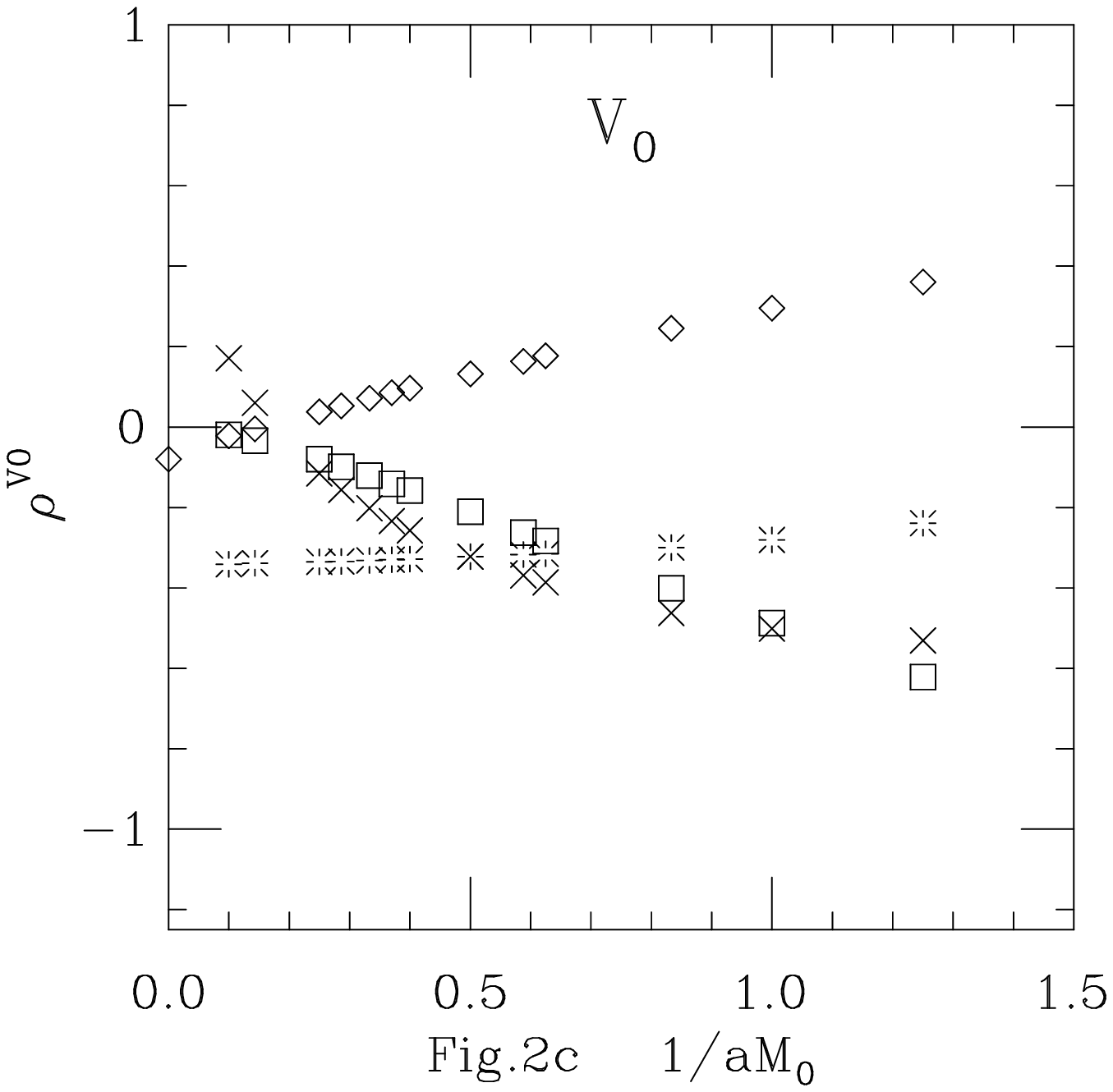}
\end{center}
\caption{ Same as Fig.1 for $A_k$, $A_0$ and $V_0$.
}
\end{figure}

\begin{figure}[t]
\begin{center}
\leavevmode
\epsfxsize=8.5cm
\epsfbox[60 120 530 430]{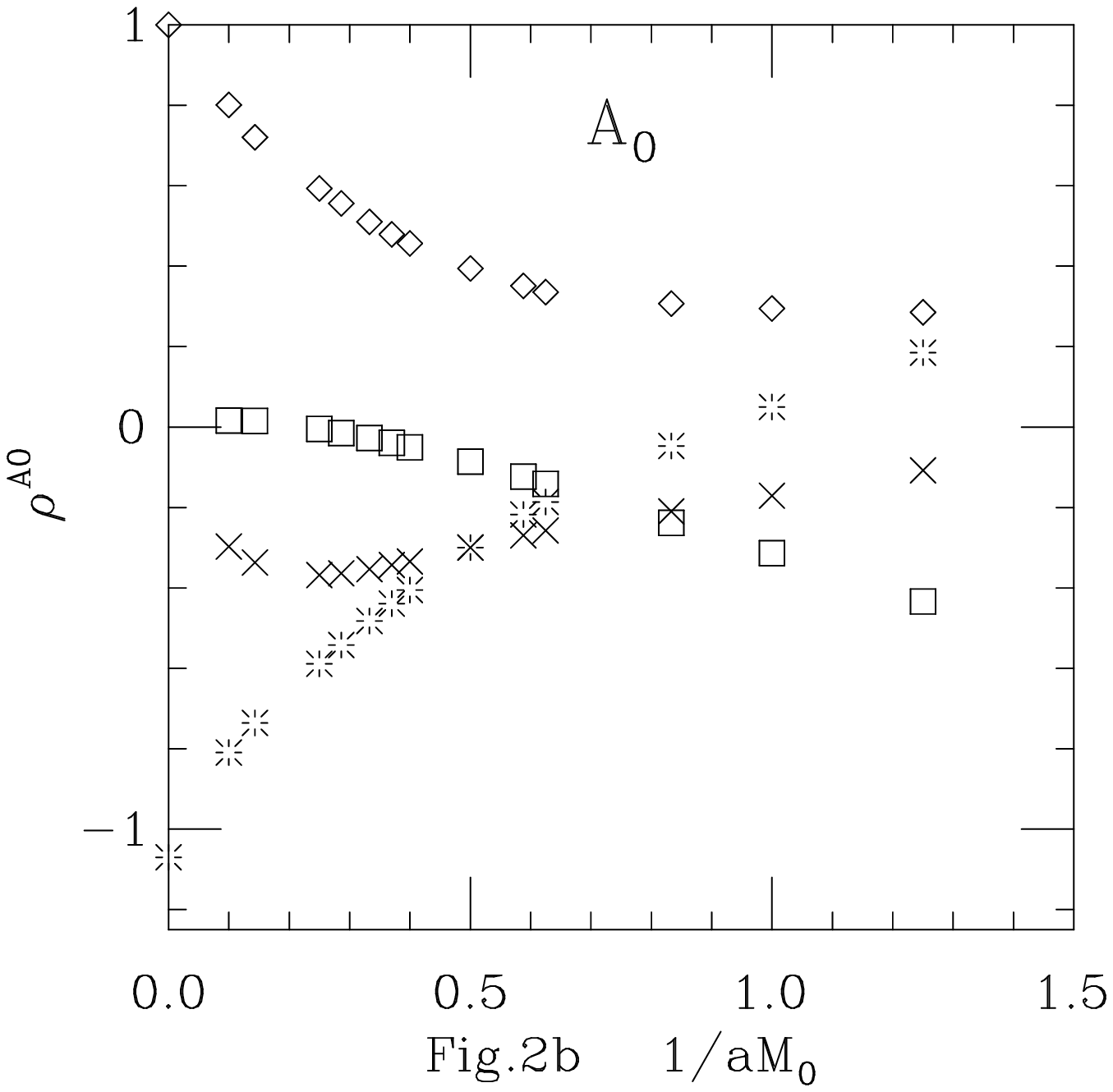}\\[25mm]
\epsfxsize=8.5cm
\epsfbox[60 120 530 430]{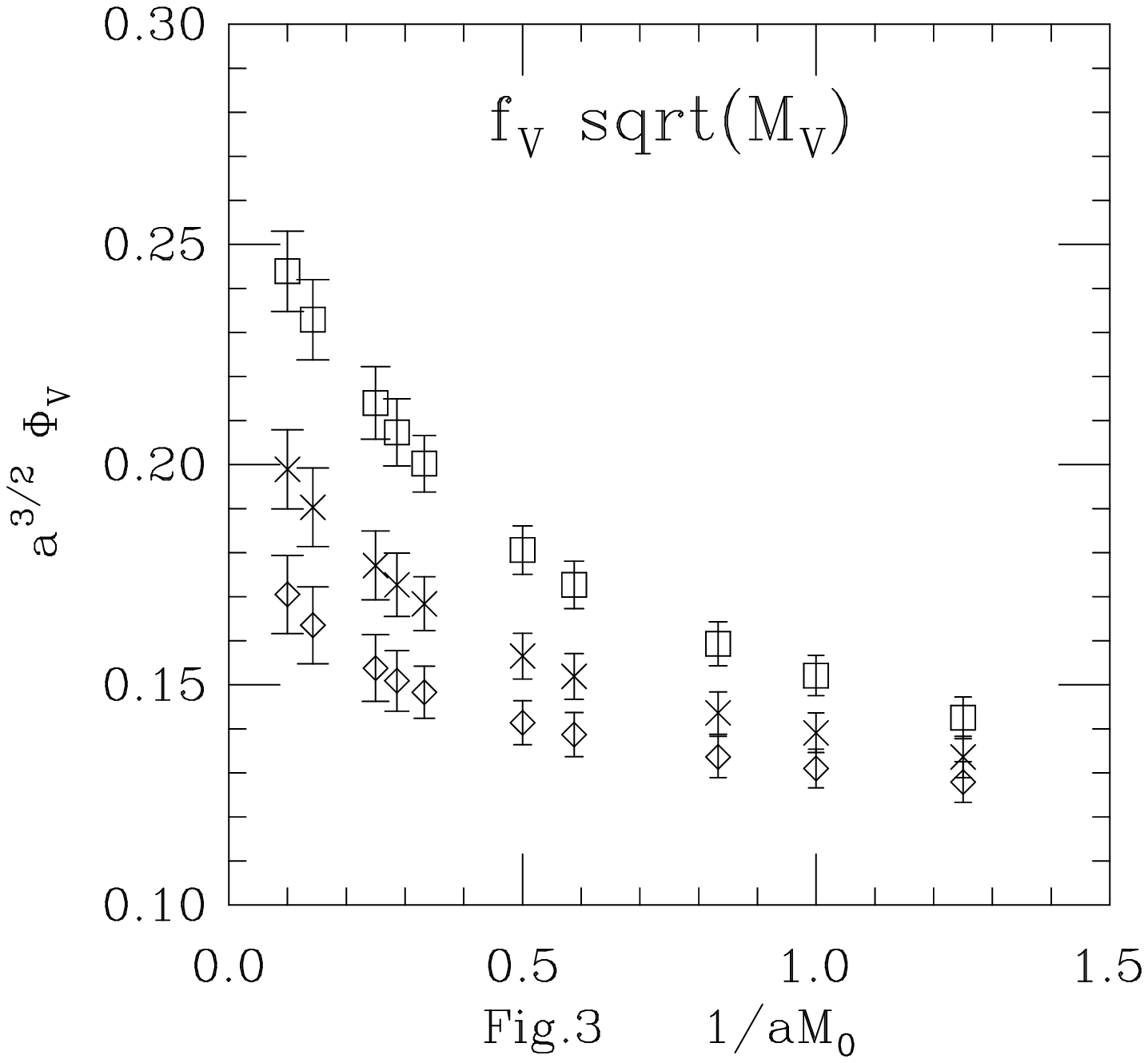}
\end{center}
\caption{ squares : tree-level ;
crosses : $\protect\alpha_V(\protect\pi/a)$ 
diamonds : $\protect\alpha_V(1/a)$.
}
\end{figure}

\section{ Applications}

Matching coefficients presented here have already been applied to
several studies of $B$ and $B^*$ meson decay constants
\cite{fb1,fb2,sara,hein}. 
In Fig.3 we show results for the vector meson decay constant 
$a^{3/2} f_V \sqrt{M_V}$ on dynamical HEMCGC configurations\cite{sara}.  
The squares represent tree-level and the crosses and diamonds 
one-loop results using $\alpha_V(q^*)$ with $q^*=\pi/a$ 
and $q^*=1/a$ respectively. The physical $B^*$ meson corresponds to 
$1/aM_0 \approx 0.5$.  The one-loop correction is a $13 \sim 21$\% 
effect depending on $q^*$.  This is a slightly larger decrease 
 than is found for 
the pseudoscalar decay constant based on $A_0$.  

\vspace{.1in}
\noindent
Acknowledgements : This work was supported by DOE Grants DE-FG03-90ER40546 
and DE-FG02-91ER40690 and by a NATO grant CRG 941259.


\begin{thebibliography}{99}


\bibitem{pert1}
For $A_0$ see also  
  C.~Morningstar and J.~Shigemitsu, Phys.\ Rev.\ D{\bf 57}, 6741 (1998).

\bibitem{bp}
A.~Borrelli and C.~Pittori; Nucl.\ Phys.\ B{\bf385}, 502 (1992).

\bibitem{fb1}
  A.~Ali Khan {\it et al.,} Phys.~Rev.~D {\bf 56}, 7012 (1997).

\bibitem{fb2}
  A.~Ali Khan {\it et al.,} Phys.\ Lett.\ B{\bf 427}, 132 (1998).

\bibitem{sara}
S.~Collins {\it et al.,} in preparation. 

\bibitem{hein}
J.~Hein {\it et al.} in preparation. 

\end{thebibliography}
\end{document}